\documentclass[aps, twocolumn, nofootinbib]{revtex4-1}
\usepackage{epsfig}
\usepackage{graphicx}
\usepackage{amsmath,amssymb,amsfonts}
\usepackage{array}
\usepackage{url}
\usepackage[usenames,dvipsnames]{color}
\usepackage[colorlinks=true,linkcolor=magenta, citecolor = blue]{hyperref}
\usepackage{multirow}
\usepackage{float}
\usepackage{lineno}
\usepackage{xspace}
\usepackage{ulem}
\usepackage{lipsum} 
\usepackage{bigints}
\begin{document}

\title{Probing Rotational Dynamics of Quark Gluon Plasma via Global Vorticity}
\author{Bhagyarathi Sahoo}
\email{Bhagyarathi.Sahoo@cern.ch}
\author{Captain R. Singh}
\email{captainriturajsingh@gmail.com}
\author{Raghunath Sahoo}
\email{Raghunath.Sahoo@cern.ch}
\affiliation{Department of Physics, Indian Institute of Technology Indore, Simrol, Indore 453552, India}

\begin{abstract}
The findings on the spin polarization of $\Lambda$, $\Xi$, and $\Omega$ hyperons and spin alignment of $K^{*0}$, $\phi$, and $D^{*+}$ mesons in relativistic heavy-ion collision experiments at the RHIC and LHC facilities propose the emergence of a strong vorticity field produced in these collisions. Contemplating the potential impact of vorticity on the space-time evolution of deconfined QCD matter and its freeze-out properties, we aim to investigate its characteristics within the medium. We introduce a complementary and data-driven approach to quantify the global vorticity field by extracting it directly from the transverse momentum spectra of produced hadrons. Employing the experimental data for $\Lambda$, $\Xi$, $\Omega$, $K^{*0}$, $K^{*\pm}$, $\phi$, $\rho$, and $D^{*+}$ at mid-rapidity in Au+Au and Pb+Pb collisions over a wide range of beam energies, $\sqrt{s_{\rm NN}}=7.7$ GeV-5.02 TeV, and centrality classes, we systematically examine spin-vorticity coupling in the medium. Our finding on the magnitude of the extracted vorticity is consistent with values deduced from $\Lambda$ and $\bar{\Lambda}$ polarization measurements using statistical thermal models under the non-relativistic limit. Notably, we observe a prominent particle-species dependence of the vorticity, as well as a non-trivial variation with collision centrality and beam energy. These results indicate that vorticity-driven spin phenomena are sensitive to hadron structure and freeze-out dynamics, providing new constraints on the rotational properties of the QCD matter.
\end{abstract}

\date{\today}
\maketitle

\section{Introduction}
\label{sec:intro}
Apprehending the nature of the fluid produced in relativistic heavy-ion collisions has been a topic of great interest over the past 
few decades, from both experimental and theoretical aspects. The large orbital angular momentum generated in non-central relativistic 
heavy-ion collisions is proposed to be partially transferred to the produced QCD medium, giving rise to a highly vortical fluid. This 
vorticity manifests itself through the global spin polarization of hyperons ($\Lambda$, $\Xi$, and $\Omega$) and the spin alignment 
of vector mesons ($K^{*0}$, $\phi$, and $D^{*+}$), as observed at RHIC and LHC energies~\cite{STAR:2017ckg, STAR:2018gyt, STAR:2019erd, 
STAR:2020xbm, ALICE:2019onw, ALICE:2021pzu, STAR:2022fan, STAR:2023eck}. The measured hyperon polarization is in qualitative 
agreement with hydrodynamic and transport model calculations that identify the vorticity field as the dominant source of spin 
polarization~\cite{Becattini:2020ngo, Li:2017slc, Fu:2020oxj, Xia:2018tes, Xie:2020}.

While the initial orbital angular momentum of the colliding nuclei provides the primary contribution to the vorticity 
field~\cite{Liang:2004ph, Jiang:2016woz}, additional sources such as inhomogeneous transverse expansion, jet-induced perturbations, 
shear viscosity, and electromagnetic fields further enhance and distort the vortical structure of the medium~\cite{Betz:2007kg, 
Xia:2018tes, Becattini:2017gcx, Pang:2016igs, Sahoo:2024egx, Sahoo:2023xnu, Einstein:1915, Sahoo:2024yud}. The resulting vorticity plays a central 
role in the phenomenology of heavy-ion collisions, affecting QCD thermodynamics~\cite{Pradhan:2023rvf}, transport 
properties~\cite{Aung:2023pjf}, phase transition~\cite{Pradhan:2023rvf}, and spin 
polarization of produced hadrons~\cite{Sahoo:2025bkx, Sahoo:2025kur, Sahoo:2025fif}. Moreover, the hydrodynamic evolution of the medium itself is 
significantly modified by vorticity~\cite{Sahoo:2023xnu}, implying that particle production at the freeze-out hypersurface retains 
direct sensitivity to the vortical structure of the expanding fireball.

Building on this sensitivity of freeze-out observables to rotational dynamics, it is essential to distinguish between different manifestations of vorticity in relativistic fluids. In heavy-ion phenomenology, vorticity is commonly divided into global and local vorticity. The global vorticity, $\boldsymbol{\Omega}_g$, characterizes the coherent, rigid-body–like rotation of the whole system, and it is associated with a velocity field of the form $\boldsymbol{v}=\boldsymbol{\Omega}_g\times\boldsymbol{R}$ for a system of characteristic size $R$. In contrast, the local vorticity, $\boldsymbol{\Omega}_l$, quantifies the microscopic circulation of the velocity field. In the non-relativistic limit, it is given by $\boldsymbol{\Omega}_l=\frac{1}{2} \boldsymbol{\nabla}\times\boldsymbol{v}$, while in relativistic hydrodynamics several inequivalent generalizations arise, including kinematic, thermal, temperature, and enthalpy vorticities, each encoding different aspects of rotational motion in a relativistic medium~\cite{Becattini:2021wqt}.

\begin{figure}[ht!]
\centering
\includegraphics[height = 6.5cm, width = 8.0cm]{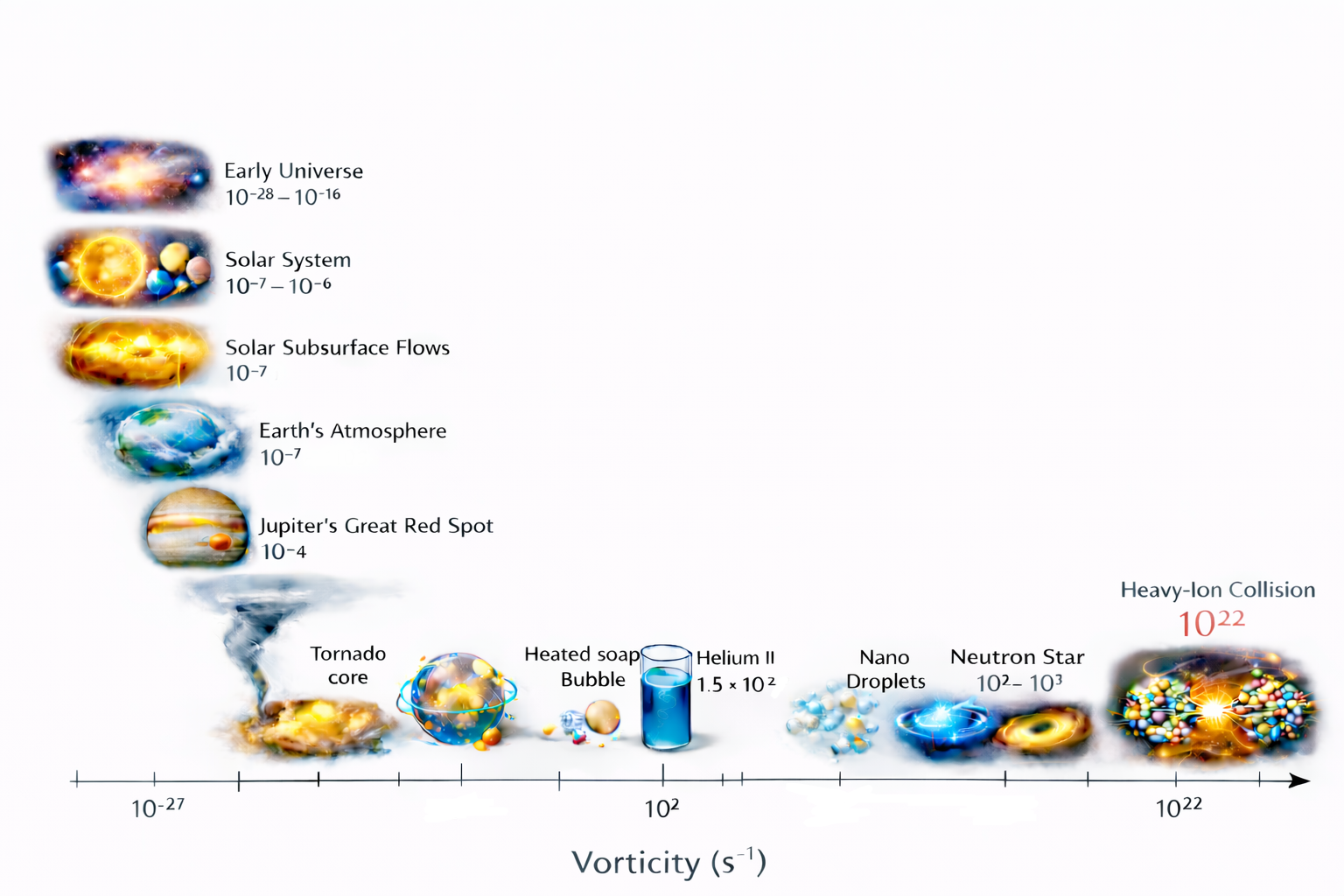}
\caption{A pictorial representation of the magnitude of the vorticity across different physical systems, ranging from the cosmos to the relativistic heavy-ion collisions laboratory.}
\label{Fig:Vor}
\end{figure}

Vorticity is a ubiquitous phenomenon across physical systems spanning many orders of magnitude in scale, ranging from heavy-ion collisions to astrophysical phenomena and objects. Examples include solar subsurface flows ($10^{-7}\;\rm s^{-1}$)~\cite{SolarSurface}, large-scale terrestrial atmospheric patterns ($10^{-7}-10^{-5}\;\rm s^{-1}$)~\cite{atmosphere}, Jupiter's Great Red Spot ($10^{-4}\;\rm s^{-1}$)~\cite{Choi:2007gpf}, supercell tornado cores ($10^{-1}\;\rm s^{-1}$)~\cite{tornado}, heated soap bubbles ($10^{2} \, \rm s^{-1}$)~\cite{soapbubble}, superfluid He II ($150 \, \rm s^{-1}$)~\cite{helium2}, and superfluid nanodroplets ($10^{7} \, \rm s^{-1}$)~\cite{nanodroplets}. Vorticity is also present in numerous astrophysical systems including the early universe ($10^{-28}-10^{-16} \, \rm s^{-1}$)~\cite{Ellis:1982xw}, solar system rotation ($10^{-7}-10^{-6} \, \rm s^{-1}$)~\cite{SolarRotation}, neutron stars ($10^{2}–10^{3} \, \rm s^{-1}$)~\cite{NeutronStar}, stellar mass black holes ($10^{3}–10^{4} \, \rm s^{-1}$)~\cite{McClintock:2006xd}, and in the laboratory in relativistic heavy-ion collisions ($10^{22} \, \rm s^{-1}$)~\cite{STAR:2017ckg}. A pictorial representation of the magnitude of the vorticity across different physical systems, ranging from the cosmos to the relativistic heavy-ion collisions laboratory, is shown in Fig.~\ref{Fig:Vor}. The magnitude of the vorticity field in heavy-ion collisions has been inferred from measurements of global $\Lambda$ and $\bar{\Lambda}$ polarization, yielding $\Omega\simeq(9\pm1)\times10^{21} \;\mathrm{s}^{-1}$~\cite{STAR:2017ckg}. More recently, complementary determinations based on directed-flow observables have been proposed~\cite{Ryu:2021lnx, Jiang:2023vxp}. A quantitative understanding of this extreme vorticity is therefore central to probing the thermodynamic, transport, and quantum properties of strongly interacting matter under rotation.

In this work, we introduce an experimentally driven, independent determination of the global vorticity by exploiting a key freeze-out 
observable: the transverse momentum ($p_{\rm T}$) spectra of hadrons. We analyze the mid-rapidity $p_{\rm T}$ distributions of single- and multi-
strange hyperons and antihyperons, $\Lambda(\bar{\Lambda})$, $\Xi^{-}(\bar{\Xi}^{+})$, and $\Omega^{-}(\bar{\Omega}^{+})$, measured in Au+Au 
collisions over a wide range of collision centralities and center-of-mass energies, $\sqrt{s_{\rm NN}}=7.7$–64 GeV. To describe these spectra, 
we employ a thermodynamically consistent Tsallis distribution formulated for a rotating medium. This framework has been shown to successfully 
reproduce hadron spectra over a broad $p_{\rm T}$ range by capturing both non-perturbative and perturbative QCD 
dynamics~\cite{Cleymans:2011in, Bhattacharyya:2015hya, Sahoo:2022iul}. We extend this formulation by incorporating rigid-body rotation, allowing the system to deviate 
slightly from local equilibrium in a physically controlled manner. The deviation from equilibrium is quantified by the non-extensive parameter $q$, while the Tsallis temperature $T$ represents an effective temperature that includes both thermal motion and collective radial flow. Further, we investigate the rotational effects of quark gluon plasma in Pb+Pb collisions at $\sqrt{s_{\rm NN}}=2.76$ TeV. We systematically studied the rotational structure of the medium at freeze-out hypersurface from RHIC to LHC energies via global vorticity\footnote{For simplicity, the global vorticity $\boldsymbol{\Omega}_g$ is denoted by $\boldsymbol{\Omega}$ hereafter}.  

Furthermore, to probe spin–vorticity coupling in the vector-meson sector, we analyze $p_{\rm T}$ spectra of $K^{*0}$ and $\phi$ mesons measured at mid-rapidity in Au+Au collisions over the center-of-mass energy range $\sqrt{s_{\rm NN}}$ = 7.7–39 GeV. We extend the analysis  to Pb+Pb collisions at $\sqrt{s_{\rm NN}} = 2.76$ and 5.02 TeV including other vector mesons, namely $K^{*\pm}$ and $\rho$. In addition, motivated by recent observations of significant spin alignment of charm vector mesons, we determine the rotation parameter $\boldsymbol{\Omega}$ for $D^{*+}$ mesons produced in Pb+Pb collisions at $\sqrt{s_{\rm NN}} = 5.02$ TeV. Although all fit parameters are obtained from the spectral analysis, the discussion primarily focuses on the centrality and beam-energy dependence of the global vorticity parameter $\boldsymbol{\Omega}$.

The remainder of this paper is organized as follows. In Section~\ref{formulation}, a detailed 
methodology is described to incorporate rotational effects into the analysis of transverse momentum spectra. Section~\ref{res} is devoted to 
the discussion of the results, with particular emphasis on the centrality and beam-energy dependence of the extracted 
vorticity. Finally, findings and their implications for vorticity-driven spin phenomena in relativistic heavy-ion collisions are 
summarized in Section~\ref{sum}.

\section{Methodology}
\label{formulation}

The experimental transverse momentum spectra of produced hadrons in $pp$ and nucleus–nucleus collisions are analyzed using the L\'{e}vy–Tsallis distribution~\cite{STAR:2006nmo, ALICE:2011gmo, ALICE:2012yqk}, which takes the following form:

\begin{align}
E \frac{d^{3}N}{dp^{3}} 
&= \frac{1}{2\pi p_{\rm T}} \frac{d^{2}N}{dp_{\rm T} dy} \nonumber \\
&= \frac{dN}{dy} \frac{(n-1)(n-2)}{2\pi n C \left[nC + m(n-2)\right]} 
\left( 1 + \frac{m_{\rm T} - m}{nC} \right)^{-n},
\label{LevyTsallis}
\end{align}

where $m_{\rm T}=\sqrt{p_{\rm T}^{2}+m^{2}}$ is the
transverse mass, $m$ is the rest mass of the particle. $\frac{dN}{dy}$, $n$ and $C$ are the fitting parameters. With this form of L\'{e}vy–Tsallis distribution, when $p_{\rm T} \gg m$, one obtains  $E \frac{d^{3}N}{dp^{3}} \propto p_{\rm T}^{-n}$, which follows a power-law distribution at high $p_{\rm T}$ accounting for the hard QCD contributions. When $p_{\rm T} \ll m$, one obtains  $E \frac{d^{3}N}{dp^{3}} \propto e^{-\frac{m_{T}-m}{C}}$, {\it i.e.}, a Boltzmannian thermal distribution.\\

An experimentally motivated improved version of the L\'{e}vy–Tsallis distribution, known as the Tsallis non-extensive distribution function, which is consistent with thermodynamic relations, is widely used to describe the $p_{\rm T}$-spectra of produced particles~\cite{Cleymans:2011in, Bhattacharyya:2015hya, Sahoo:2022iul}, 
 \begin{equation}
 \frac{d^{2}N}{dp_{\rm T}dy} = \frac{g Vp_{\rm T} E}{(2\pi)^{2}}
 \left[1+(q-1)\frac{(E-\mu)}{T}\right]^{-\frac{q}{q-1}}
 \label{spectra}
 \end{equation}
\noindent where $g$ denotes the degeneracy factor, $V$ is the system volume. At mid rapidity, i.e., $y$ = 0,  $E = m_{\rm T}=\sqrt{p_{\rm T}^{2}+m^{2}}$ is the single-particle energy (in natural units). Notably, the vacuum hadron mass $m$ is used in the single particle energy relation. In Eq. ~\ref{spectra}, $\mu$ is the chemical potential. At LHC energies, $\mu$ is set to zero, while at RHIC energies its value is taken from Ref.~\cite{Andronic:2017pug}. The parameter $T$ in Eq.~\ref{spectra} incorporates both random thermal motion and collective flow effects. The non-extensive parameter $q$, often referred to as the entropy index~\cite{Tsallis}, quantifies the deviation of the system from local thermodynamic equilibrium and accounts for the transition from exponential behavior at low $p_{\rm T}$ to a power-law tail at high $p_{\rm T}$. Equation~\ref{LevyTsallis} is connected to Eq.~\ref{spectra} by the transformation $n \rightarrow q/(q-1)$ and $nC \rightarrow \frac{T+m (q-1)}{q-1}$~\cite{Cleymans:2013rfq}.

A thermodynamic system undergoing rigid rotation with angular velocity $\boldsymbol{\Omega}$ responds mechanically through its total angular momentum $\boldsymbol{J}$, which is the thermodynamic conjugate variable to $\boldsymbol{\Omega}$. For systems composed of particles with finite spin, such as hyperons and vector mesons, $\boldsymbol{J}$ contains both orbital and intrinsic spin contributions. The presence of rotation modifies the single-particle energy measured in the inertial laboratory frame, $E^{\rm (lab)}$, leading to the relation~\cite{Landau};
 \begin{equation}
 E = E^{\rm (lab)} - \boldsymbol{J}\cdot\boldsymbol{\Omega}
 \label{eq2}
 \end{equation}
\noindent here, Eq.~\ref{eq2} can be derived from the single-particle Hamiltonian in a rotating frame~\cite{Mashhoon:1988zz, Hehl:1990nf}. Further, Eq.~\eqref{eq2} demonstrates that vorticity induces a shift in the single-particle energy through spin–orbit coupling. Incorporating this energy modification into the Tsallis distribution of Eq.~\eqref{spectra} enables the extraction of the global vorticity from $p_{\rm T}$-spectra measured in relativistic heavy-ion collisions. Such global vorticity serves as a valuable probe for studying the response of the quark-gluon plasma to local vorticity in most of the analytical and numerical investigations of its thermodynamic properties. Therefore, understanding the behavior of the system under global rotation is equally important while studying the spin polarization of hyperons and the spin alignment of vector mesons arising from local vortical structures.\\

\section{Results and Discussion}
\label{res}
\begin{figure*}[ht!]
\centering
\includegraphics[scale = 0.29]{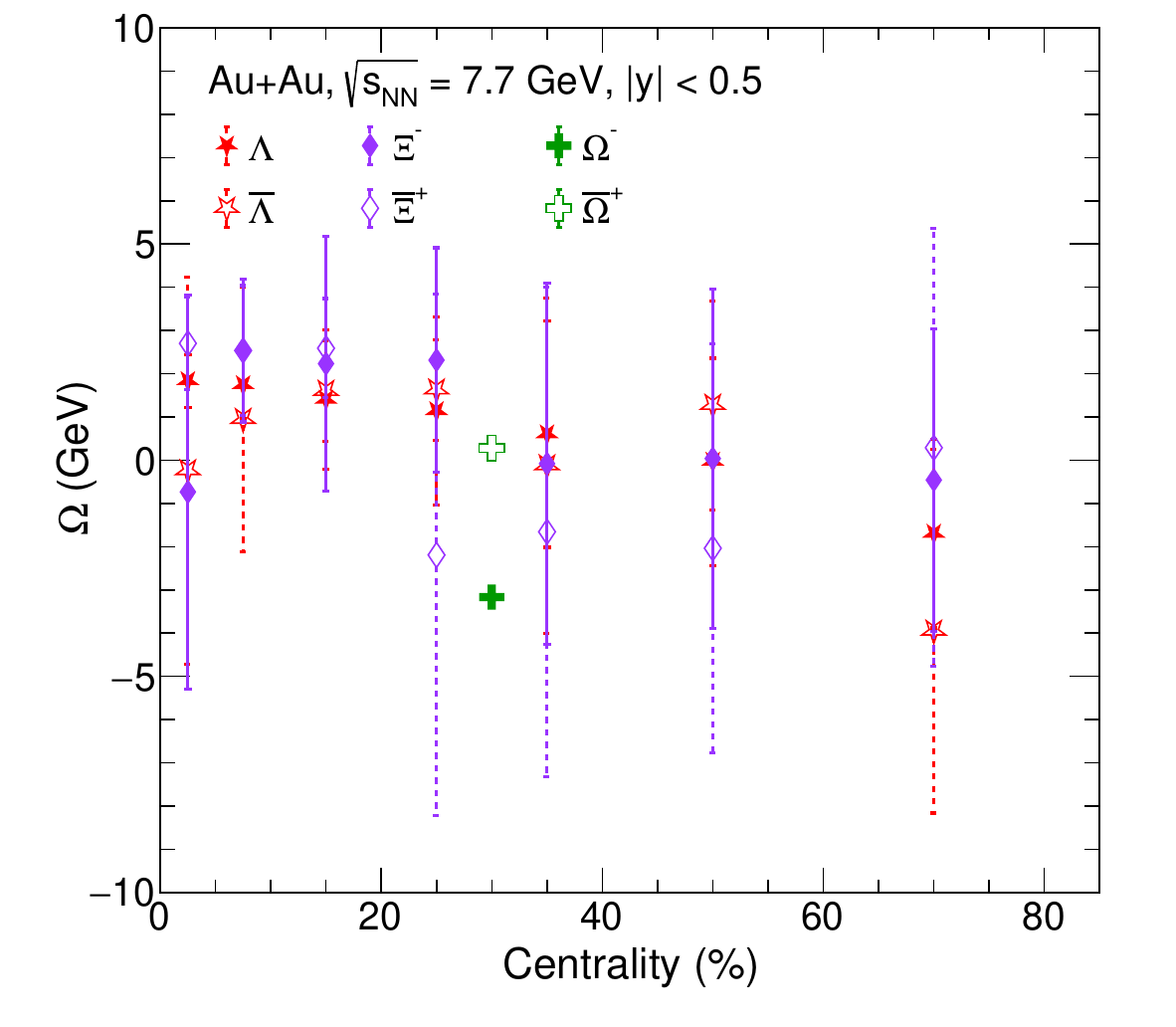}
\includegraphics[scale = 0.29]{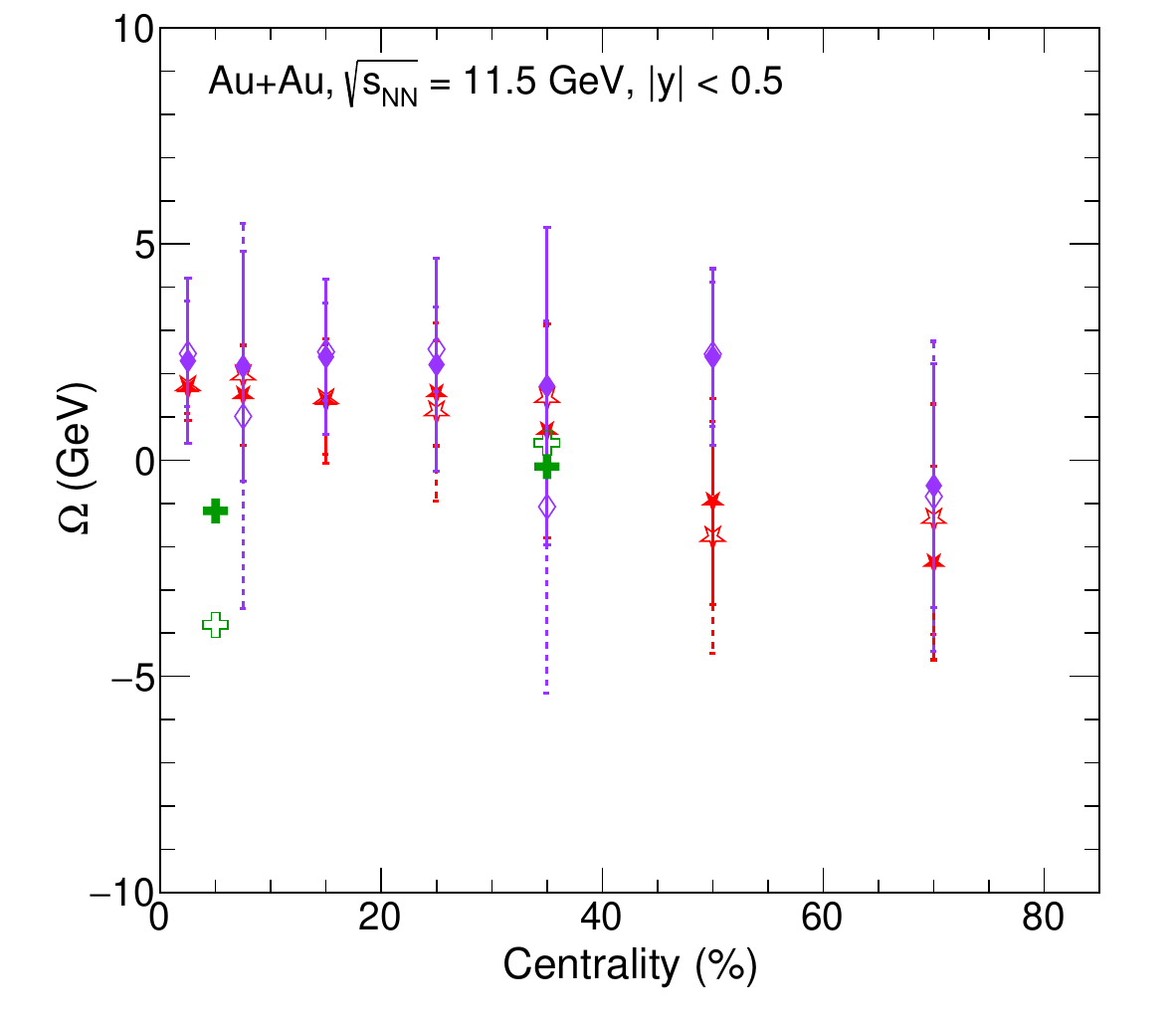}
\includegraphics[scale = 0.29]{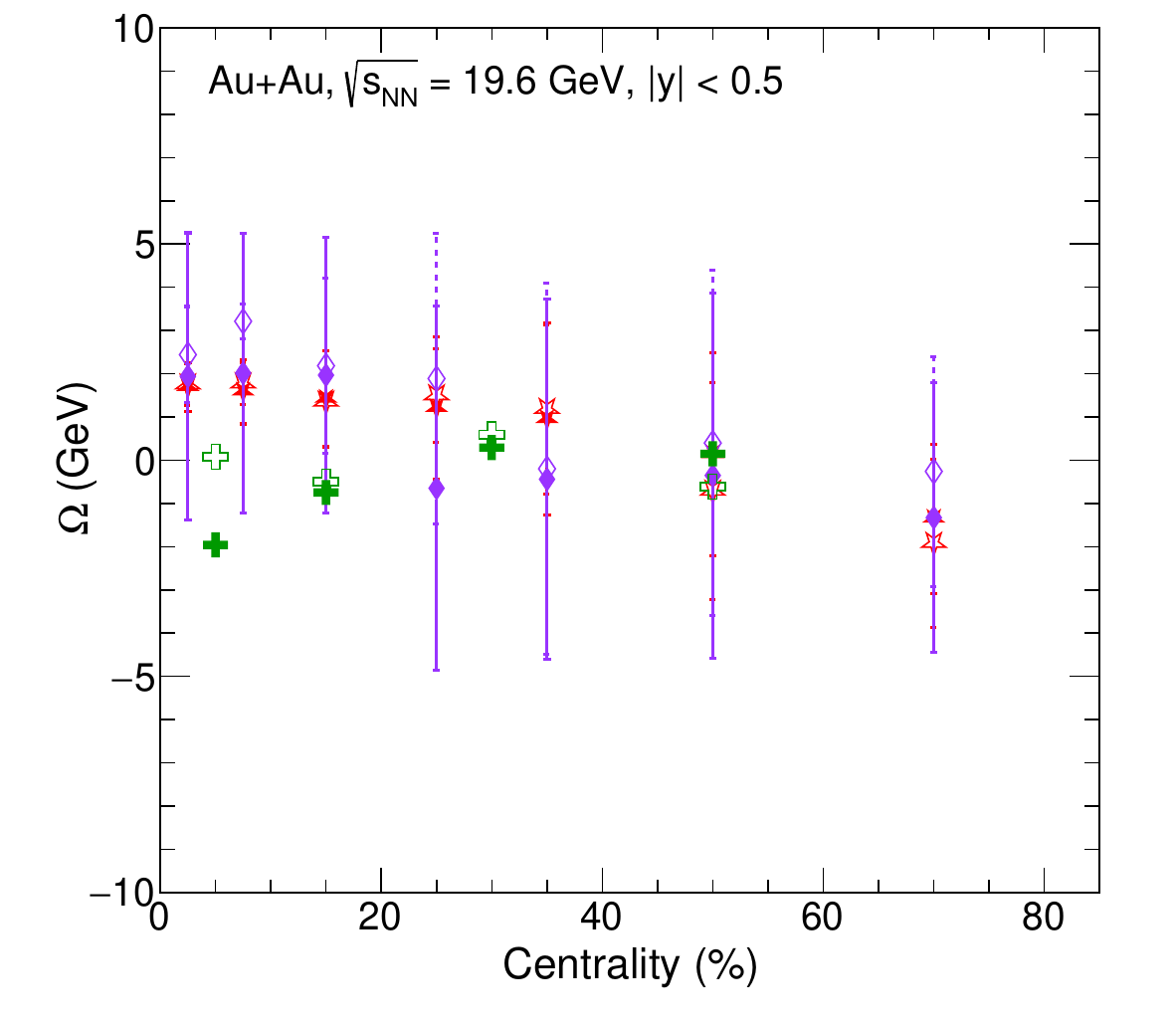}
\includegraphics[scale = 0.29]{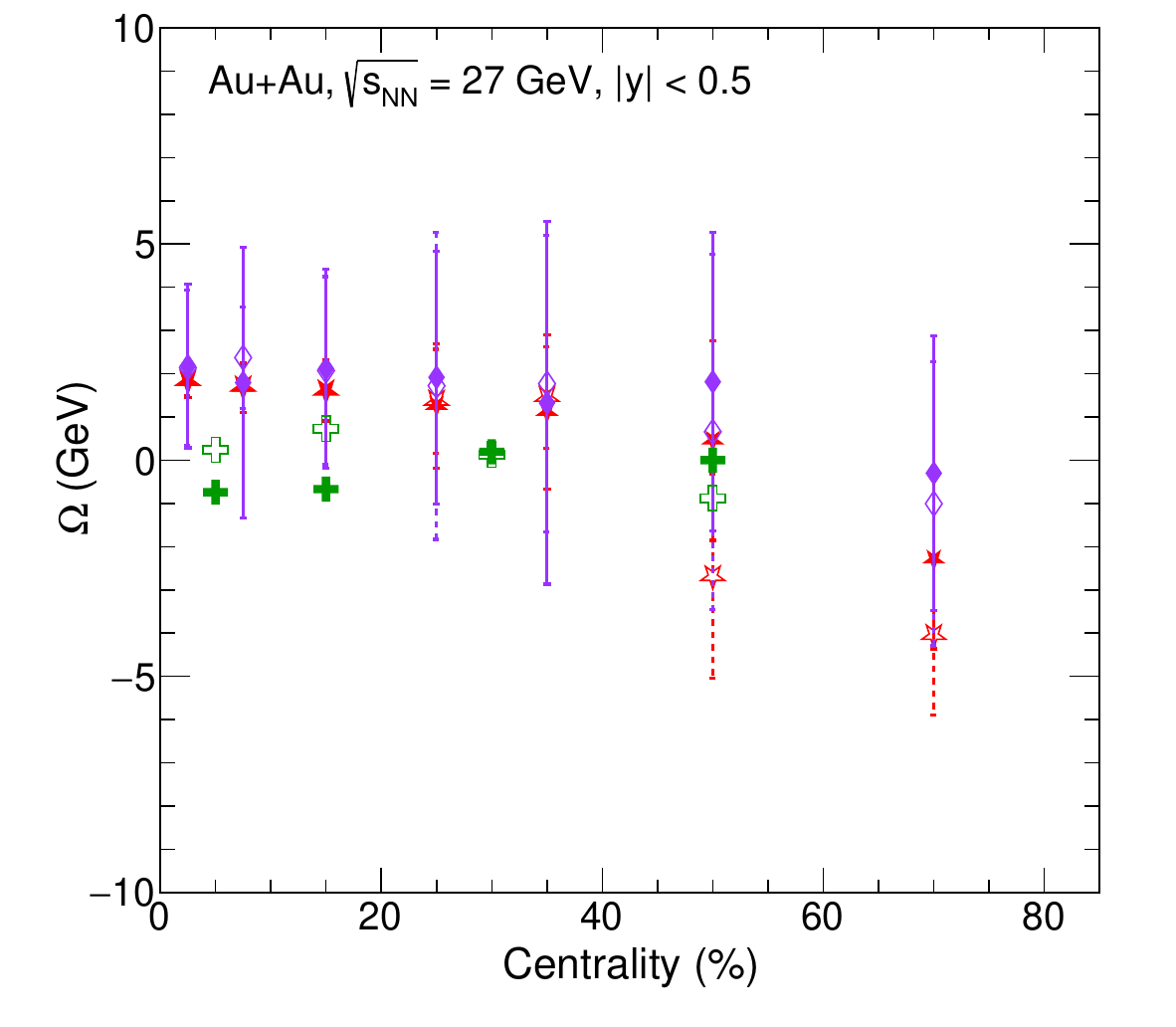}
\includegraphics[scale = 0.29]{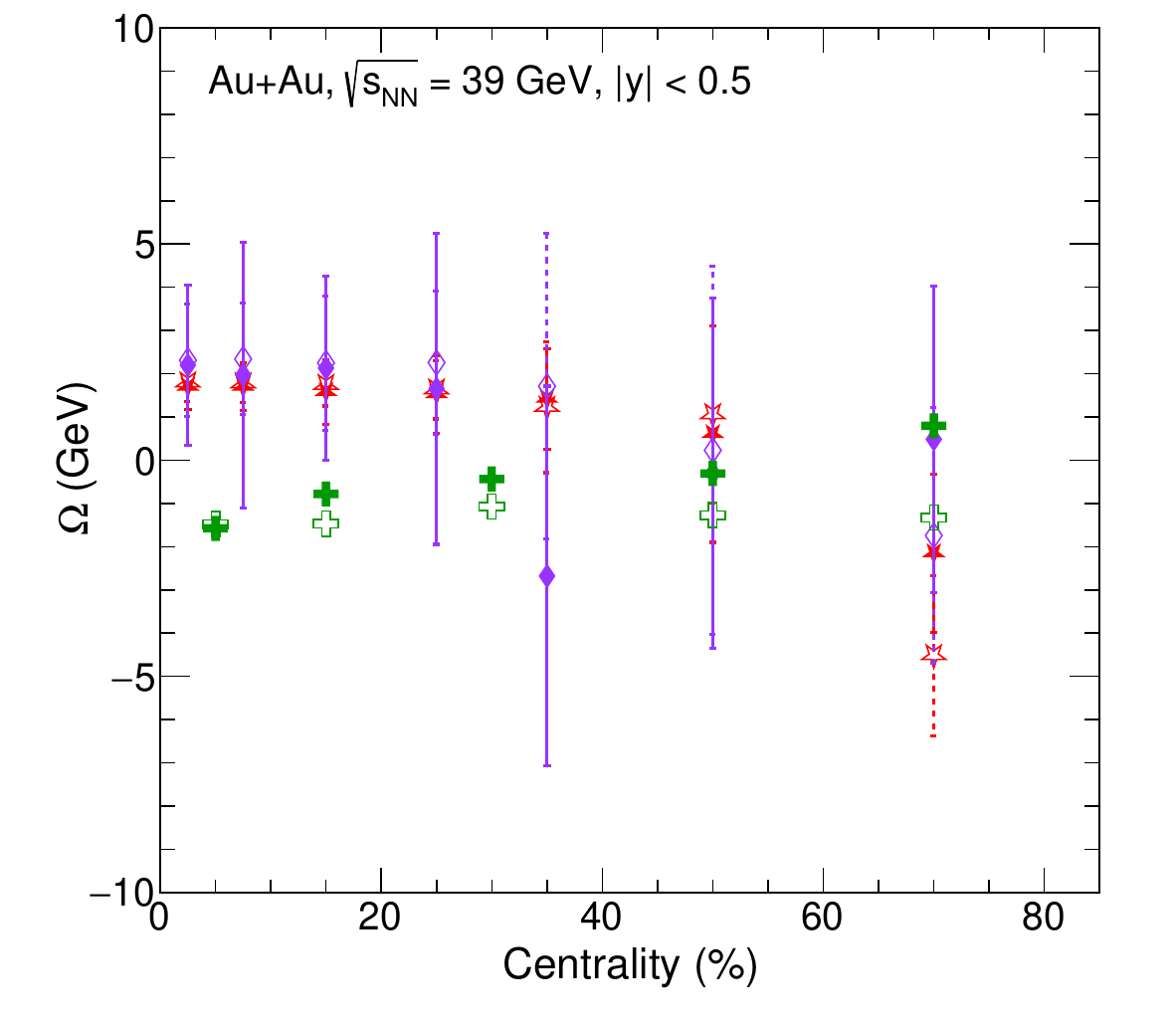}
\includegraphics[scale = 0.29]{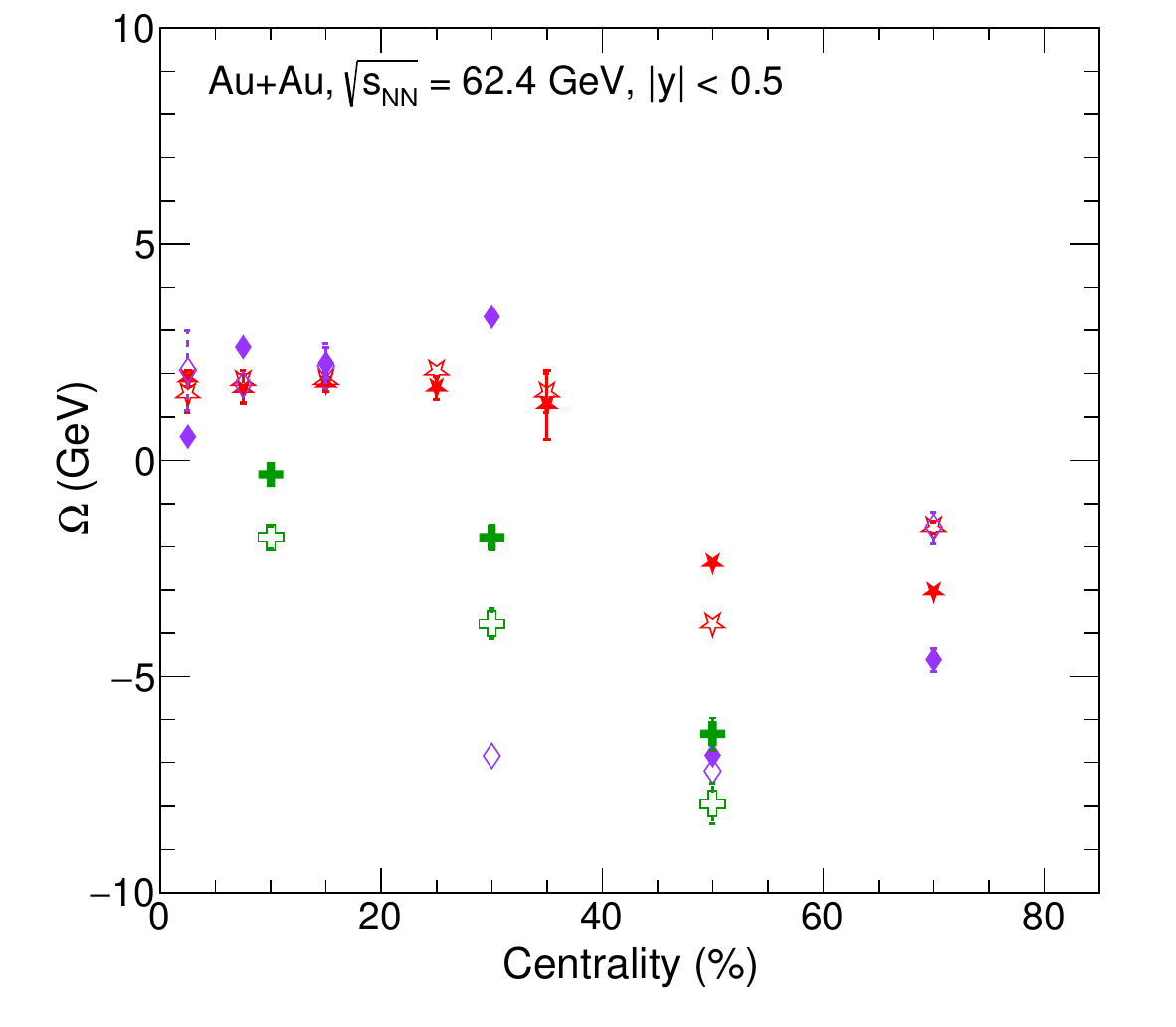}
\caption{The global vorticity $\Omega$ as a function of collision centrality for $\Lambda$, $\bar{\Lambda}$, $\Xi^{-}$, $\bar{\Xi}^{+}$, $\Omega^{-}$, and $\bar{\Omega}^{+}$ obtained from Au+Au collisions  at mid-rapidity for various center of mass energies ranging from $\sqrt{s_{\rm NN}}$ = 7.7–64 GeV.}
\label{Fig:RotvsCenthyperons}
\end{figure*}
The centrality dependence of the global vorticity $\Omega$, extracted from experimentally measured $p_{\rm T}$-spectra of various hadrons by the STAR and ALICE collaborations, is presented in this section. The analysis first focuses on single- and multi-strange hyperons, examining the variation of $\Omega$ with collision centrality and center-of-mass energy, and is then extended to vector mesons. These results demonstrate how spin–vorticity coupling can be accessed at the freeze-out hypersurface.

Notably, the best-fit values of the Tsallis parameters
(and their uncertainties) along with the $\chi^2/ndf$ are provided in a supplemental files for all considered hadrons and center of mass energies, attached to the original manuscript. Further, the comparison of the Tsallis parameters, i.e., effective temperature $T$, non-extensive parameter $q$, freeze-out volume $V$ for a non-rotating baseline ($\Omega$=0) with the rigid rotation case  ($\Omega \neq$0) are also presented in the supplemental files. The $\Omega$ values presented in this study are determined using the maximum transverse momentum fit range accessible in the corresponding experimental data.

\subsection{Strange Baryons}
\subsubsection{Global vorticity at RHIC energies}
Figure~\ref{Fig:RotvsCenthyperons} shows the centrality dependence of the extracted global vorticity $\Omega$ for $\Lambda$, $\bar{\Lambda}$, $\Xi^{-}$, $\bar{\Xi}^{+}$, $\Omega^{-}$, and $\bar{\Omega}^{+}$ hyperons produced in Au+Au collisions at mid-rapidity at $\sqrt{s_{\rm NN}}=7.7$–64 GeV. The experimental data used in the present study are taken from Ref.~\cite{STAR:2019bjj, STAR:2010yyv}. For $\Lambda$ and $\bar{\Lambda}$ hyperons, $\Omega$ decreases toward peripheral collisions across most beam energies, remaining positive in central and mid-central collisions and becoming negative in peripheral events. The sign of $\Omega$ reflects the direction of the rotation of the system, with positive (negative) values corresponding to clockwise (anticlockwise) rotation. A similar centrality dependence is observed for the doubly strange $\Xi^{-}$ and $\bar{\Xi}^{+}$ hyperons. In contrast, the triply strange $\Omega^{-}$ and $\bar{\Omega}^{+}$ hyperons exhibit a qualitatively different behavior. At intermediate energies ($\sqrt{s_{\rm NN}}=11.5$–39 GeV), $\Omega$ increases toward peripheral collisions, whereas at $\sqrt{s_{\rm NN}}=62.4$ GeV the trend reverses. This non-uniform centrality dependence suggests that multi-strange hyperons respond differently to the vortical medium, likely reflecting their larger mass, higher strangeness content, and earlier kinetic freeze-out.

For a given centrality class, the extracted vorticity differs among $\Lambda$, $\Xi^{-}$, and $\Omega^{-}$ hyperons, indicating a clear particle-species dependence of spin–vorticity coupling at freeze-out. This differential response arises from the distinct intrinsic properties of the hyperons, including mass, spin, and quark content, which influence their coupling to the rotating medium and shape their transverse momentum spectra. Notably, the magnitude of $\Omega$ obtained from this spectral analysis is consistent, within uncertainties, with values inferred from $\Lambda$ and $\bar{\Lambda}$ spin polarization measurements using statistical thermal models in the non-relativistic limit~\cite{STAR:2017ckg}.
\subsubsection{Global vorticity at LHC energies}
 \begin{figure}[ht!]
\centering
\includegraphics[height = 6.0cm, width = 7.2cm]{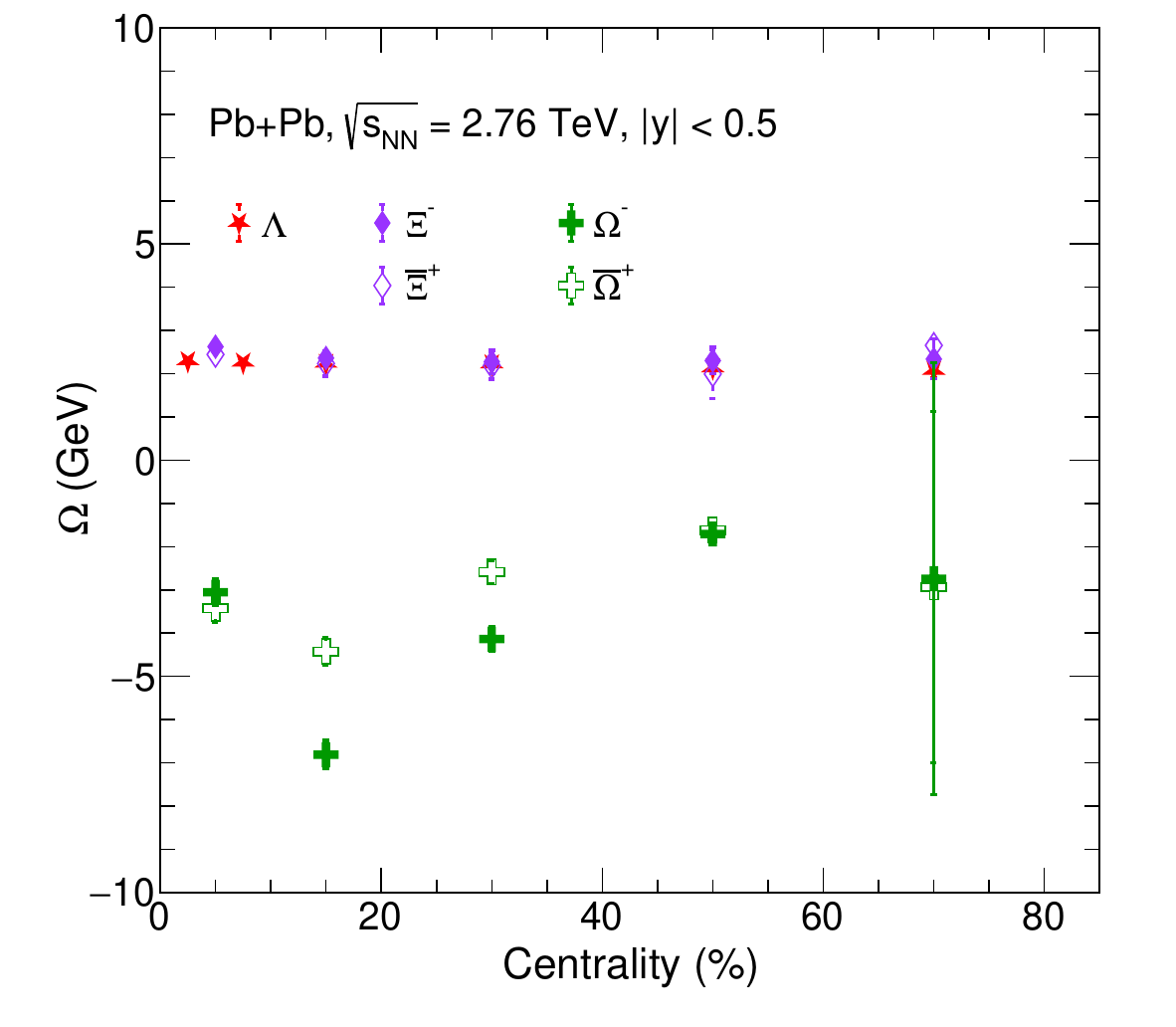}
\caption{The global vorticity $\Omega$ as a function of collision centrality for $\Lambda$,  $\Xi^{-}$, $\bar{\Xi}^{+}$, $\Omega^{-}$, and $\bar{\Omega}^{+}$ obtained in Pb+Pb collisions at mid-rapidity ($|y| <$ 0.5) for $\sqrt{s_{\rm NN}} =$ 2.76 TeV.}
\label{Fig:PbPbCenthyperons}
\end{figure}

 \begin{figure}[ht!]
\centering
\includegraphics[scale = 0.35]{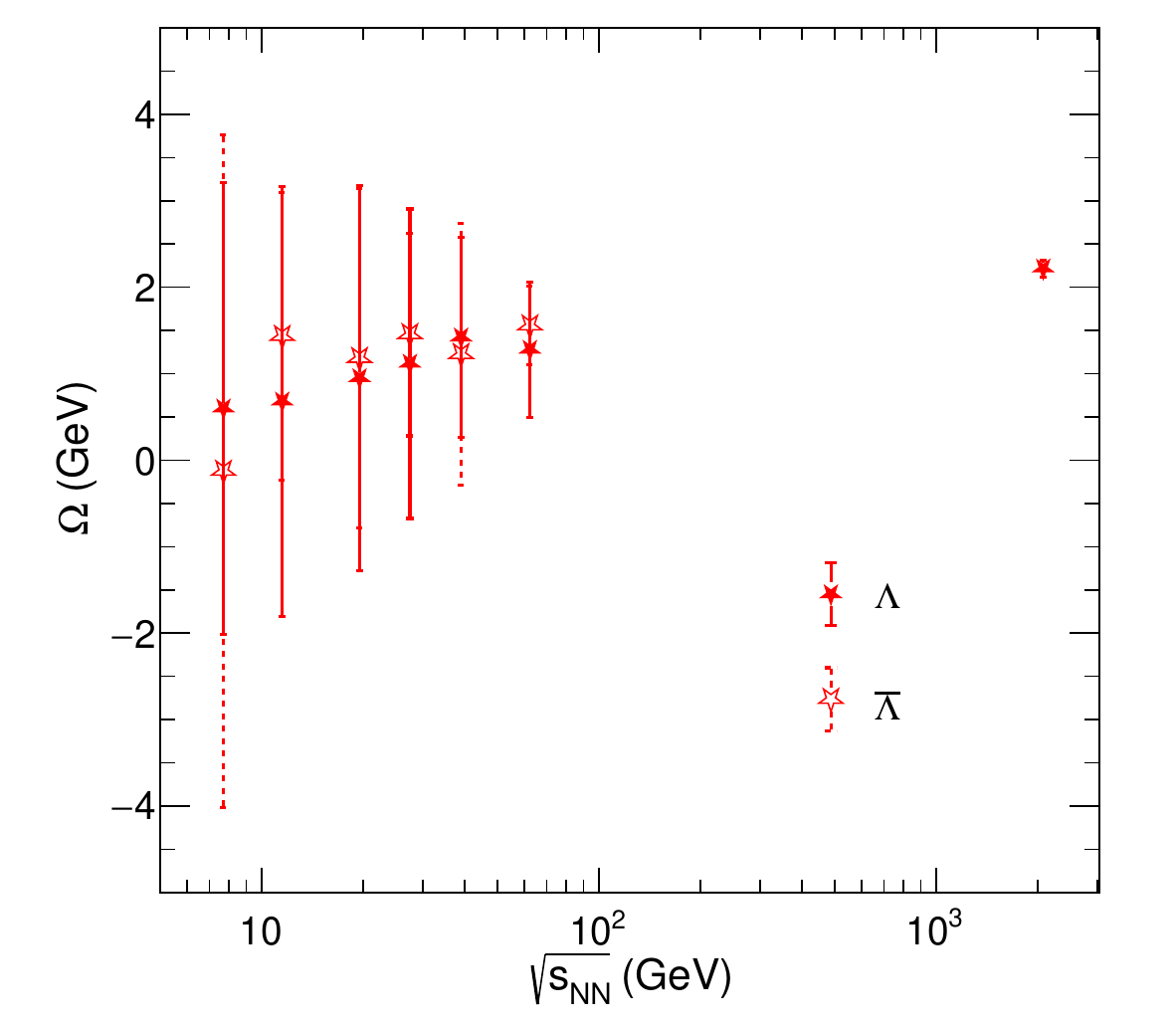}
\caption{The global vorticity $\Omega$ as a function of collision center of mass energies $\sqrt{s_{\rm NN}}$ for $\Lambda$, and $\bar{\Lambda}$ obtained in Au+Au and Pb+Pb collisions at mid-rapidity ($|y| <$ 0.5) for 30-40\% and 20-40\% centrality classes, respectively.}
\label{Fig:RotvsSNN}
\end{figure}
At LHC energies, experimental data used for the analysis are taken from Ref.~\cite{ALICE:2013xmt}. Figure~\ref{Fig:PbPbCenthyperons} illustrates the centrality dependence of $\Omega$ for $\Lambda$,  $\Xi^{-}$, $\bar{\Xi}^{+}$, $\Omega^{-}$, and $\bar{\Omega}^{+}$ hyperons produced in Pb+Pb collisions at $\sqrt{s_{\rm NN}}=2.76$ TeV at mid rapidity. For $\Lambda$ and $\Xi^{-}$ hyperons, the centrality dependence of $\Omega$ is significantly weakened, remaining approximately flat up to 80\% centrality. This behavior may indicate modifications in the relaxation dynamics of vortical structures or changes in baryon transport at higher energies. In contrast, the $\Omega^{-}$ hyperon exhibits a pronounced centrality dependence, highlighting the enhanced sensitivity of multi-strange baryons to rotational effects. These trends raise questions about how vorticity evolves as the collision energy increases and how the medium redistributes its total angular momentum between orbital and spin angular momentum. The behavior of $\Omega$ is found to be similar for $\Xi^{-}$ and $\bar{\Xi}^{+}$ for all centrality classes, while a mild difference in  $\Omega$ is observed for $\Omega^{-}$ and $\bar{\Omega}^{+}$ for (10-40)\% at LHC energy.

It is important to note that the extracted $\Omega$  primarily reflects hadronic freeze-out dynamics rather than being a direct measure of QGP vorticity. However, this does not preclude a meaningful connection to QGP rotational dynamics. In relativistic heavy-ion collisions at RHIC and LHC energies, the properties of the partonic matter are inferred indirectly from freeze-out observables recorded at the detector level. The transverse momentum spectra of single- and multi-strange hadrons carry information about the ongoing partonic processes during hadronization, and the collective behavior of the produced particles is widely understood to be a consequence of the expansion of the QGP medium. The effective temperature $T$ extracted from the Tsallis fit carries contributions from both the random thermal motion and the collective radial flow, and is related to the kinetic freeze-out temperature $T_\mathrm{fro}$ via the relation $T = T_\mathrm{fro} + m\langle u_{t} \rangle^{2}$. Where $\langle u_{t} \rangle$ is the average radial transverse flow velocity. The measured $\Omega$, obtained from the analysis of the transverse momentum spectra, can therefore be interpreted as an effective rotational parameter that constrains the QGP rotational dynamics to some extent. This interpretation is analogous to how the radial flow velocity and the freeze-out temperature are widely used to infer bulk QGP properties from hadronic spectra, even though these quantities are themselves measured at the hadronic level. We further note that the strange and multi-strange hadrons ($\Lambda$, $\Xi$, $\Omega$) used in this analysis are particularly well-suited for this purpose. Due to their small hadronic interaction cross-sections, they decouple earlier from the hadronic medium and are therefore less affected by late-stage hadronic rescattering. Their freeze-out conditions are thus expected to be closer to the partonic phase, making the extracted $\Omega$ a more reliable proxy for QGP rotational dynamics. Additionally, rigid rotation can effectively influence the evolution of the viscous QGP medium by modifying the velocity profiles, as explicitly reported in Ref.~\cite{Florkowski:2017ruc}, and demonstrated in Ref.~\cite{Sahoo:2023xnu}.

The center-of-mass energy dependence of the global vorticity $\Omega$ is shown in Fig.~\ref{Fig:RotvsSNN}, which presents $\Omega$ as a function of $\sqrt{s_{\rm NN}}$ for $\Lambda$ and $\bar{\Lambda}$ hyperons measured at mid-rapidity ($|y|<0.5$) in Au+Au and Pb+Pb collisions for the 30–40\% and 20–40\% centrality classes, respectively. The extracted vorticity is found to be significantly larger at LHC energies than at RHIC energies. This enhancement reflects the larger initial orbital angular momentum $\boldsymbol{L}$ deposited in the system at higher beam energies, where $\boldsymbol{L}$ depends on the beam momentum $\boldsymbol{p}_{\rm beam}$ and impact parameter $\boldsymbol{b}$ through $\boldsymbol{L}\simeq\boldsymbol{p}_{\rm beam}\times\boldsymbol{b}$. Since $\boldsymbol{p}_{\rm beam}$ increases with $\sqrt{s_{\rm NN}}$, the initial orbital angular momentum scales approximately linearly with the collision energy. The results indicate that converting a fraction of the initial orbital angular momentum into global vorticity becomes more efficient at higher energies, reaching a maximum at the LHC. In a rigidly rotating system, the angular momentum is related to the angular velocity through $L=I\Omega$, where $I$ is the moment of inertia. Thus, it is expected to have the higher global vorticity at LHC energies compared to RHIC energies. However, this picture can be different for a fluid local vorticity case in a hydrodynamically evolving relativistic system, where the thermal expansion of the medium strongly depends on the gradients of the hydrodynamic field variables such as velocity field and temperatures, and enthalpy, etc. Under such conditions, the conversion of initial orbital angular momentum into local and global vorticity can be highly non-linear and strongly dependent on the dynamics of the medium evolution.

Furthermore, from Fig.~\ref{Fig:RotvsSNN} it is evident that the global vorticity generated in relativistic heavy-ion collisions induces the same rotational alignment in hyperons and anti-hyperons. This observation indicates that the spin-vorticity coupling acts identically on particles and antiparticles, unlike the magnetic-field-induced Zeeman coupling, which leads to opposite spin alignments for particles and antiparticles due to their opposite magnetic moments. However, the strength of the spin–vorticity coupling is found to depend on particle mass. Here, it is important to mention that, although the vorticity field originates from the collective conversion of the initial orbital angular momentum of the system, its manifestation in rotational alignment varies among hadrons. Heavier particles couple more strongly to the vortical field and therefore exhibit a more pronounced rotational alignment than lighter ones. This mass-dependent behavior closely parallels the observed particle-species dependence of anisotropic flow coefficients, such as directed, elliptic, and triangular flow, in which a common collective background yields different magnitudes depending on hadron mass, quark content, and hadronization dynamics. Similarly, in the presence of a shared vortical medium, the resulting rotational alignment reflects the intrinsic properties of the produced hadrons.
\begin{figure*}[ht!]
\centering
\includegraphics[scale = 0.25]{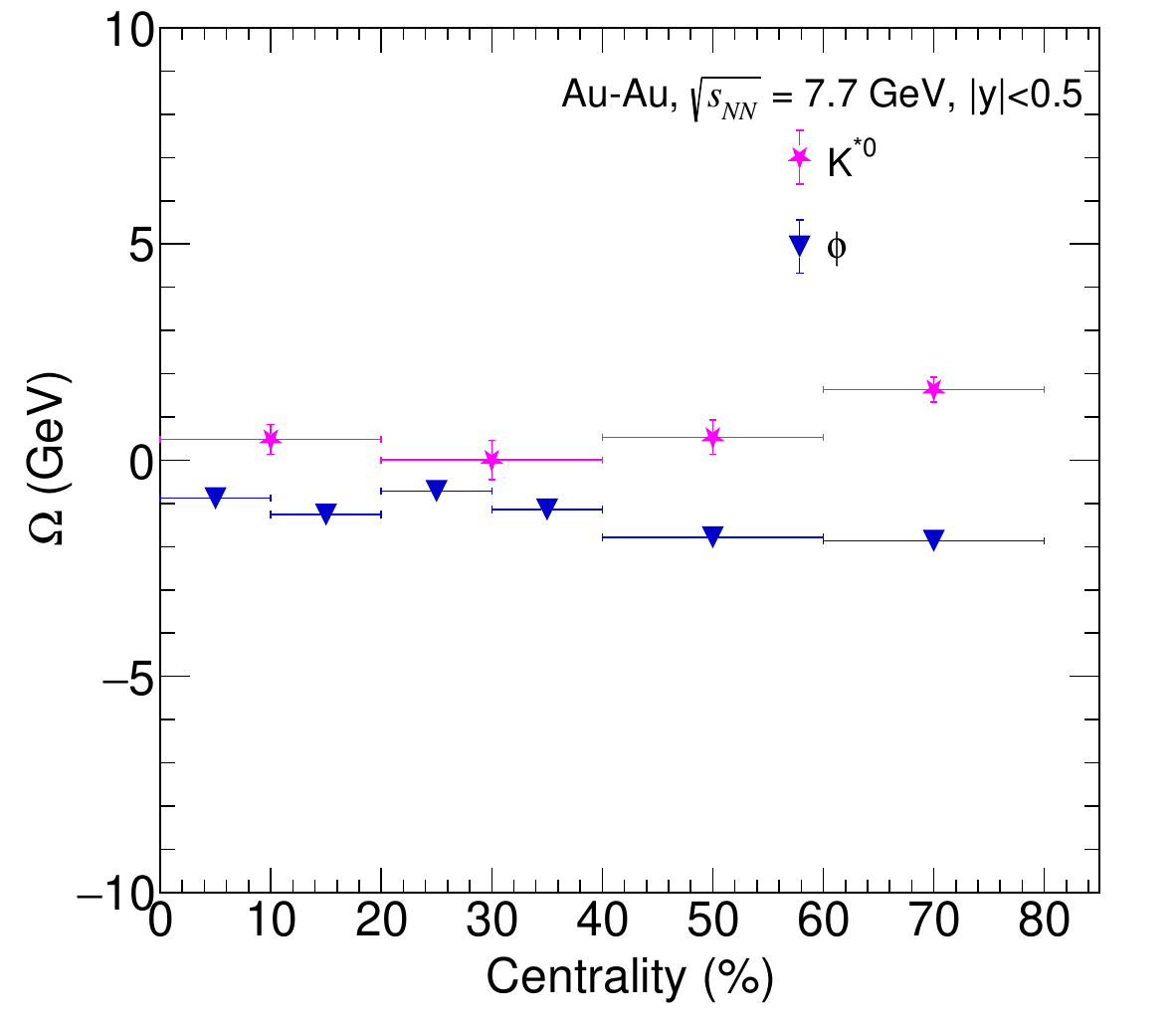}
\includegraphics[scale = 0.25]{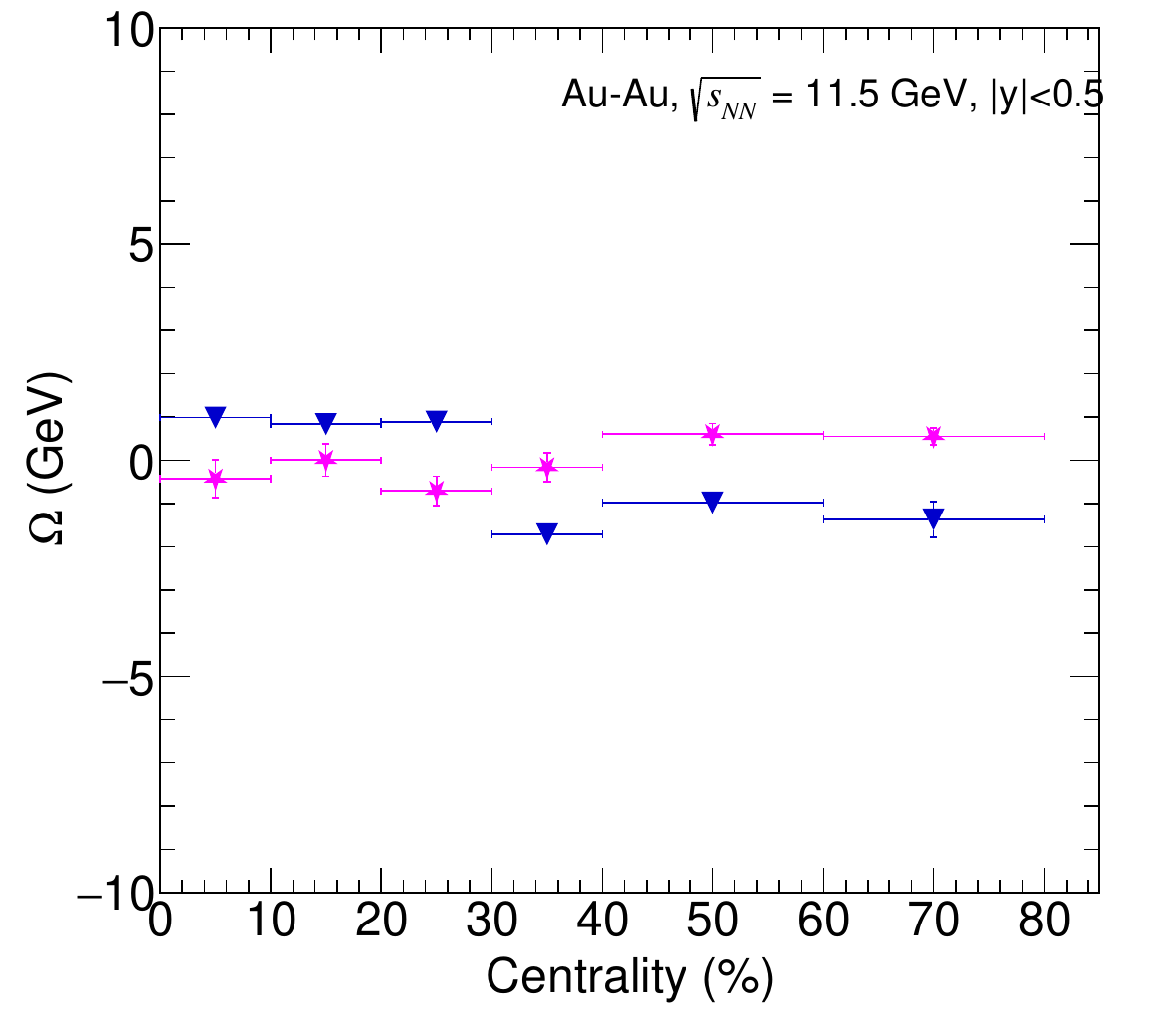}
\includegraphics[scale = 0.25]{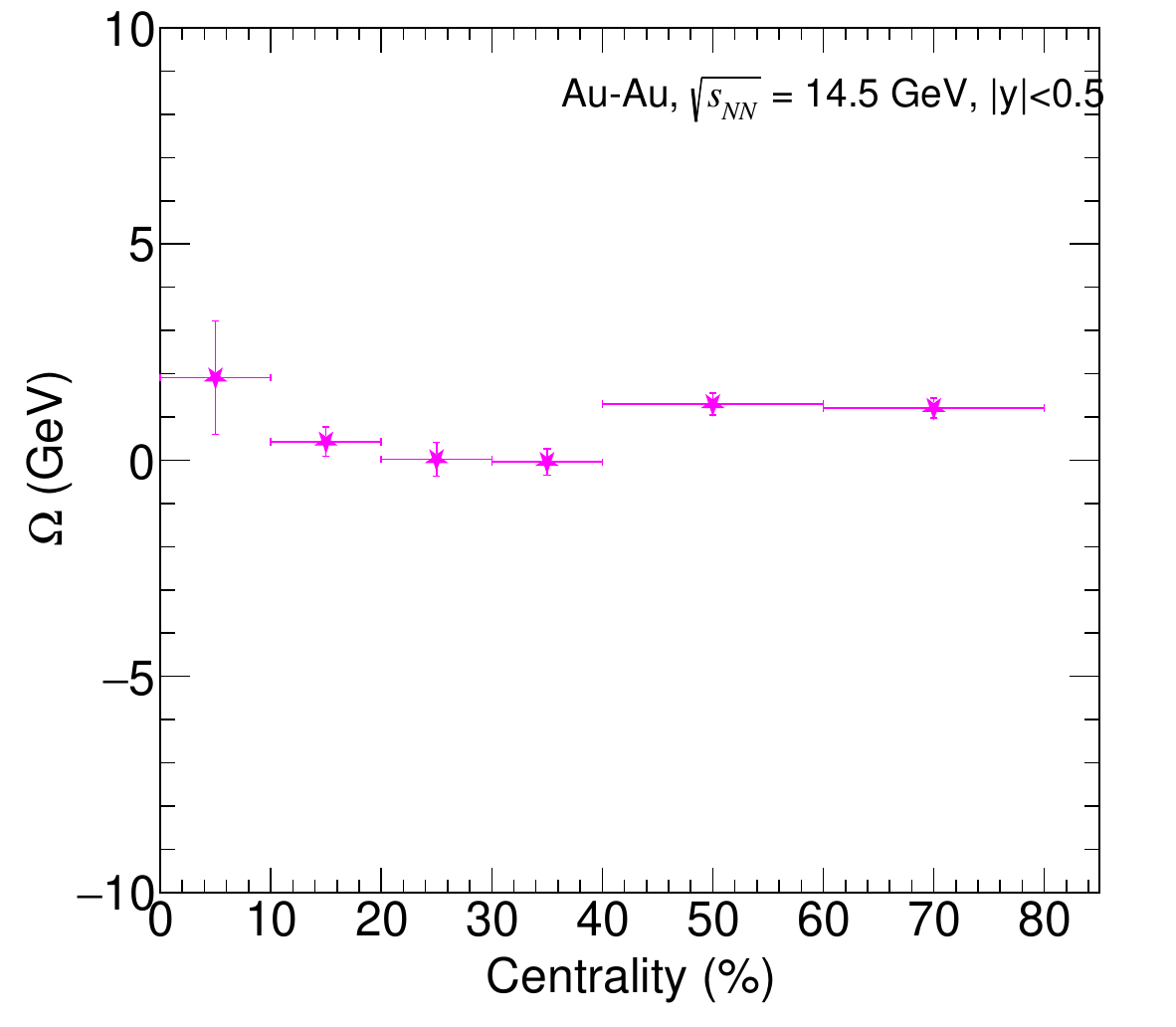}
\includegraphics[scale = 0.25]{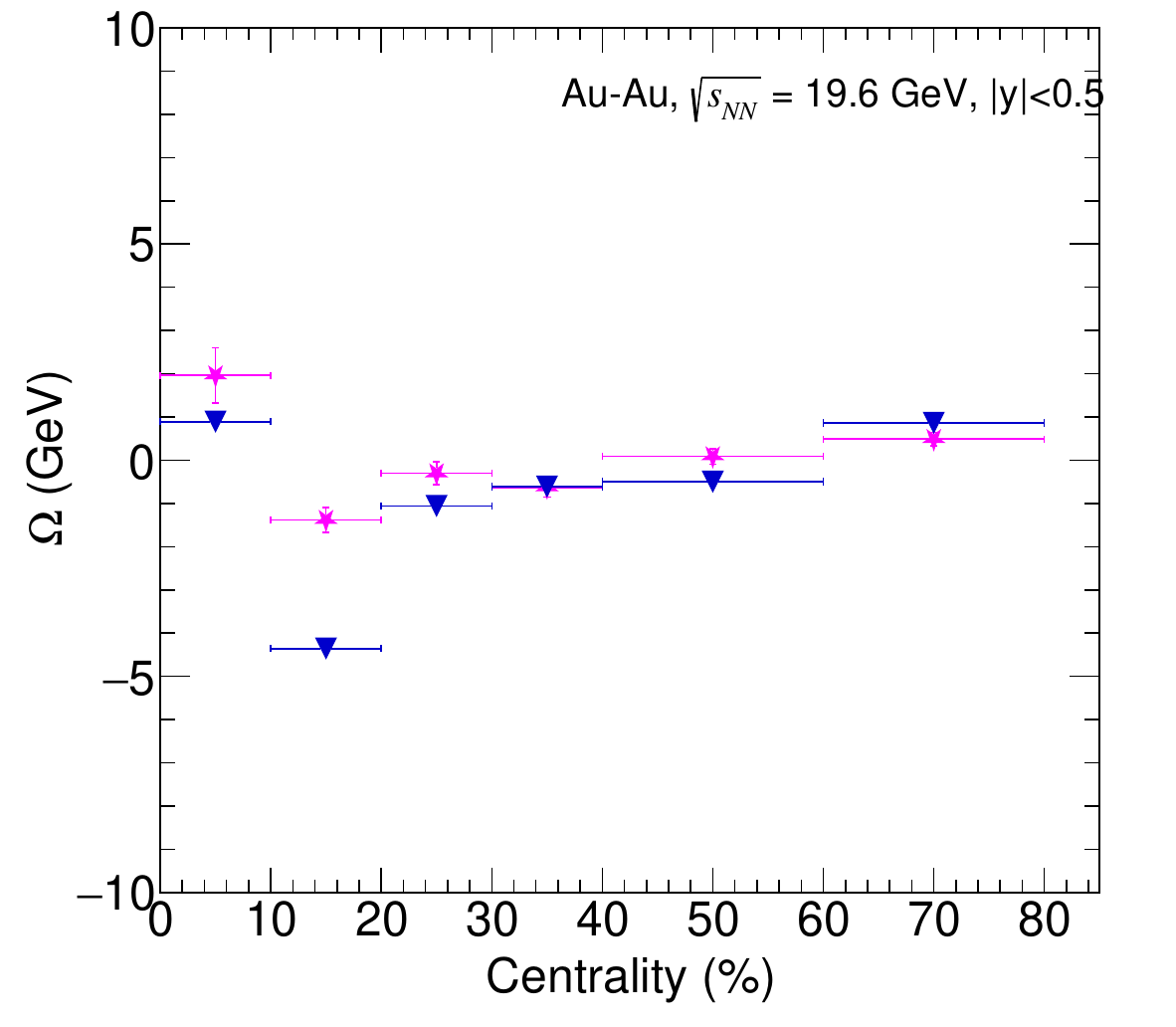}
\includegraphics[scale = 0.25]{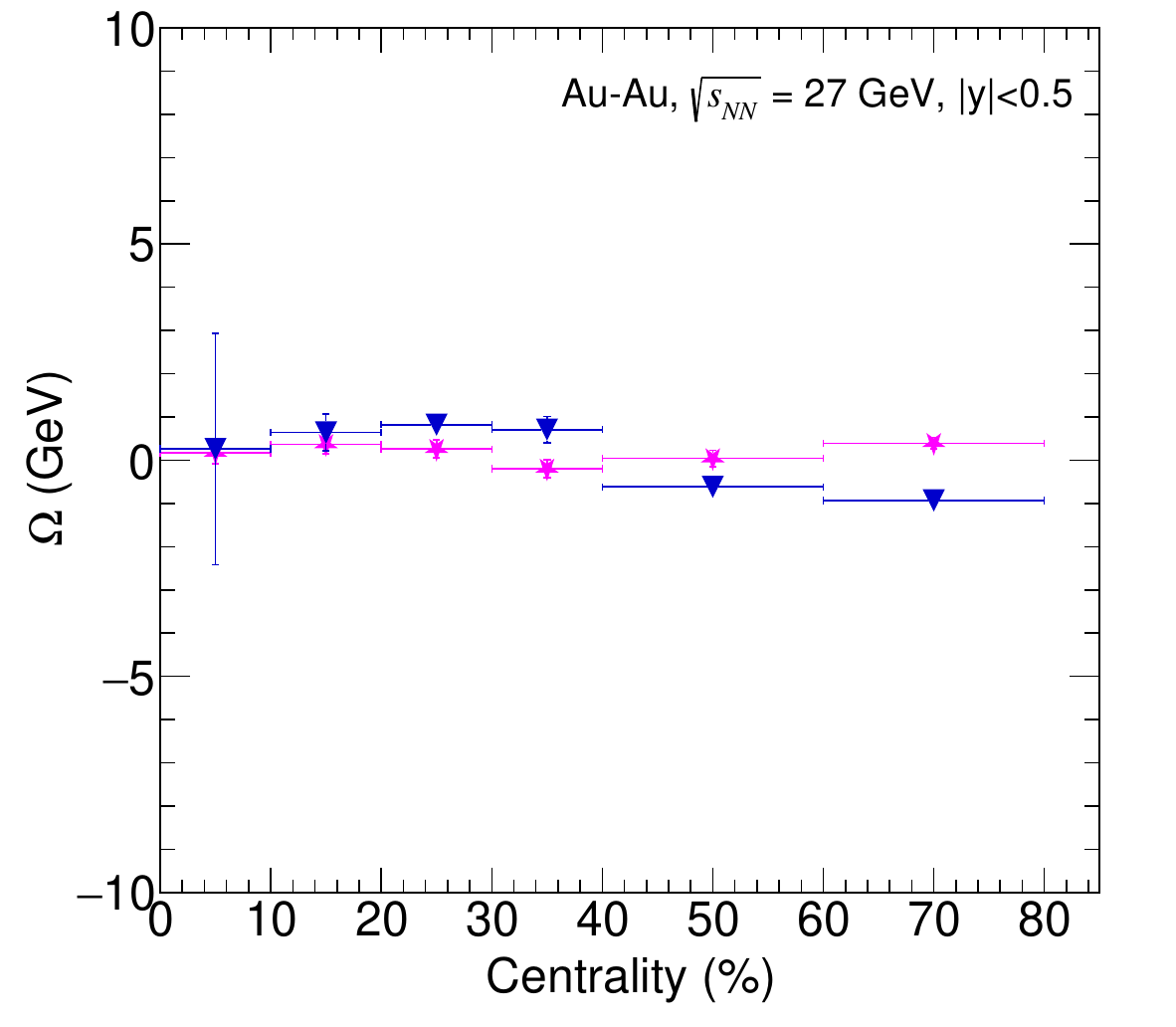}
\includegraphics[scale = 0.25]{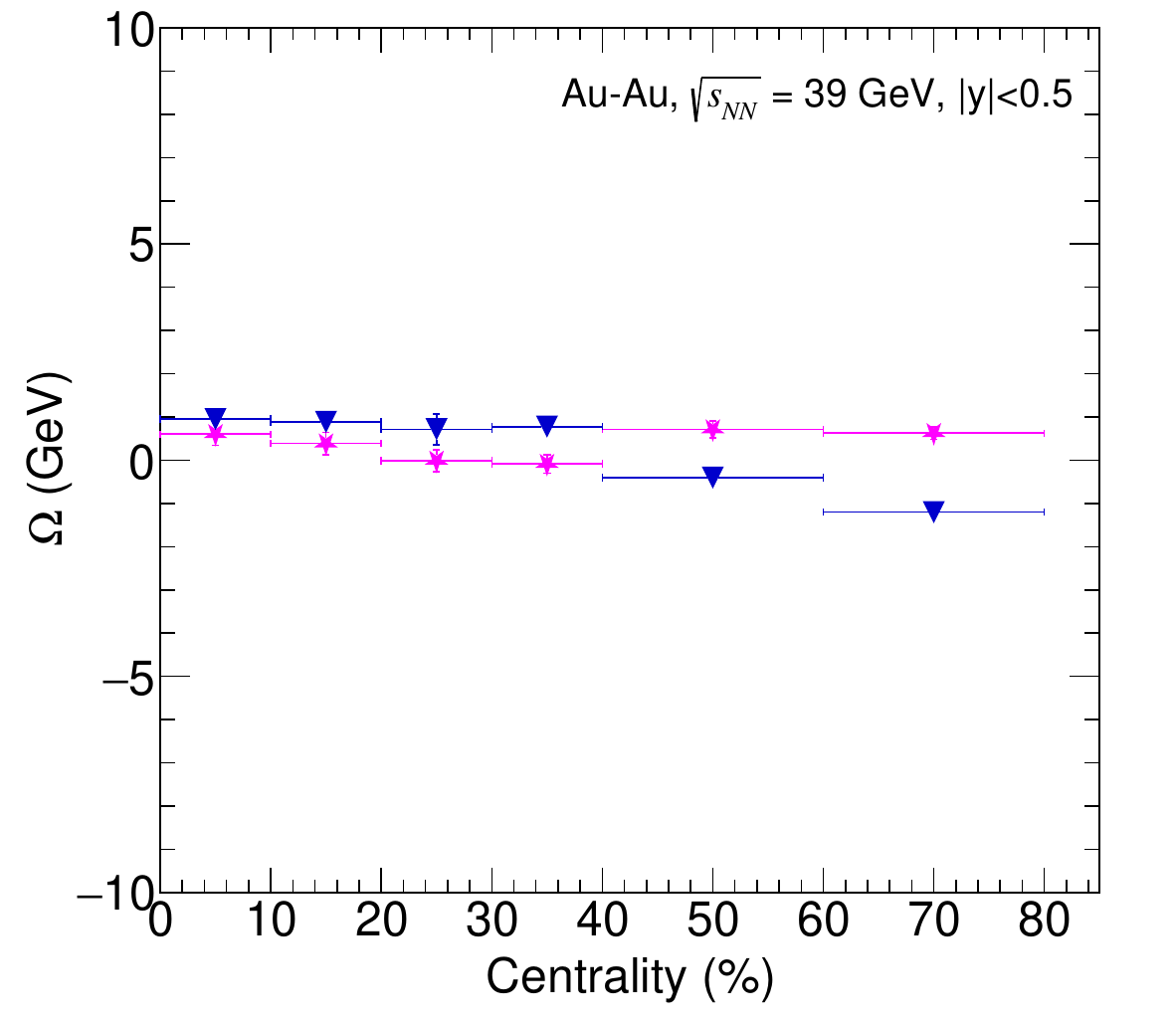}
\caption{The global vorticity $\Omega$ for $K^{*0}$, and $\phi$ mesons as a function of collision centrality obtained from Au+Au collisions  at mid-rapidity for various center of mass energies ranging from $\sqrt{s_{\rm NN}}$ = 7.7–39 GeV.}
\label{Fig:STARMesons}
\end{figure*}
\begin{figure*}[ht!]
\centering
\includegraphics[height = 5.8cm, width = 7.2cm]{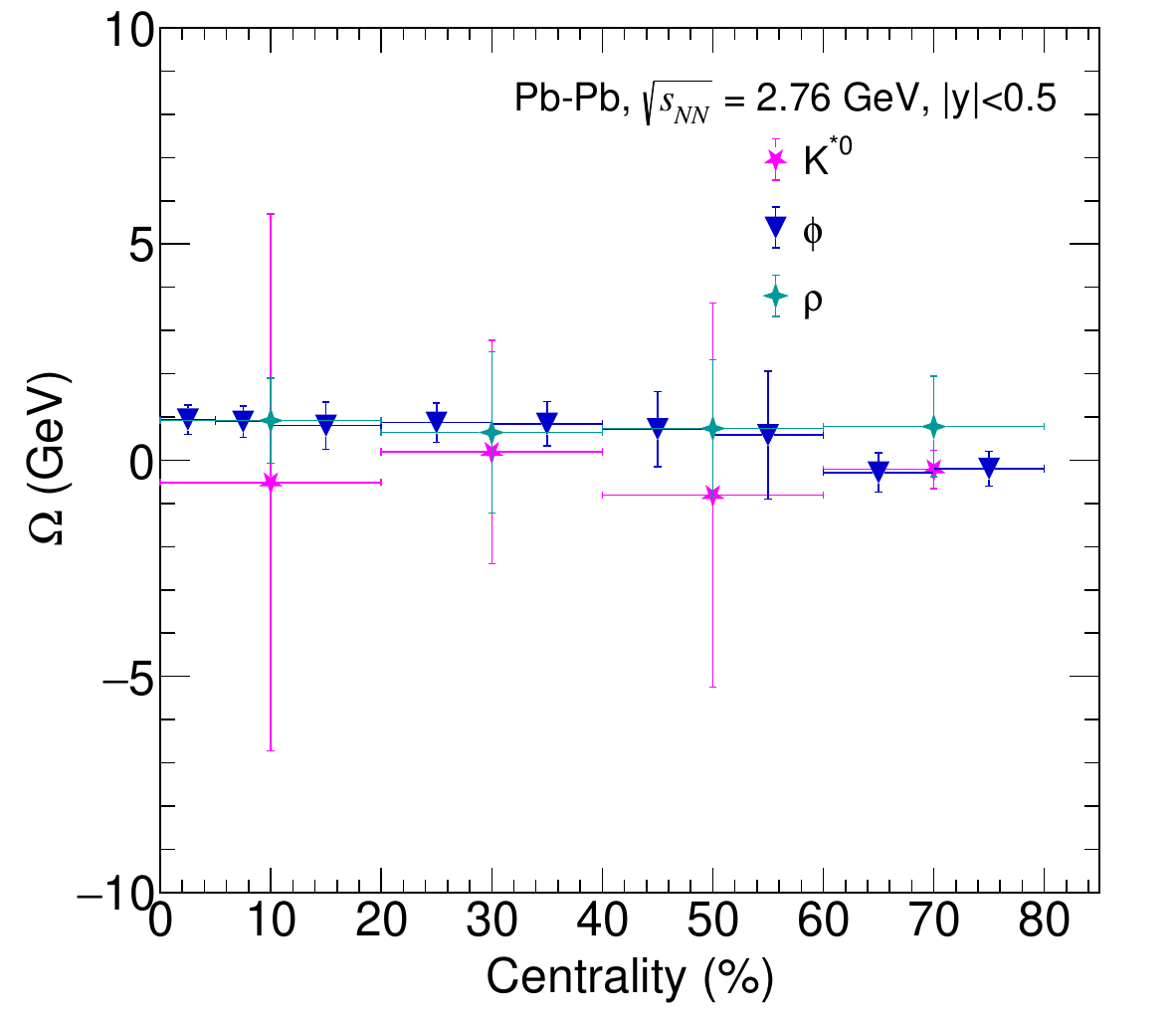}
\includegraphics[height = 5.8cm, width = 7.2cm]{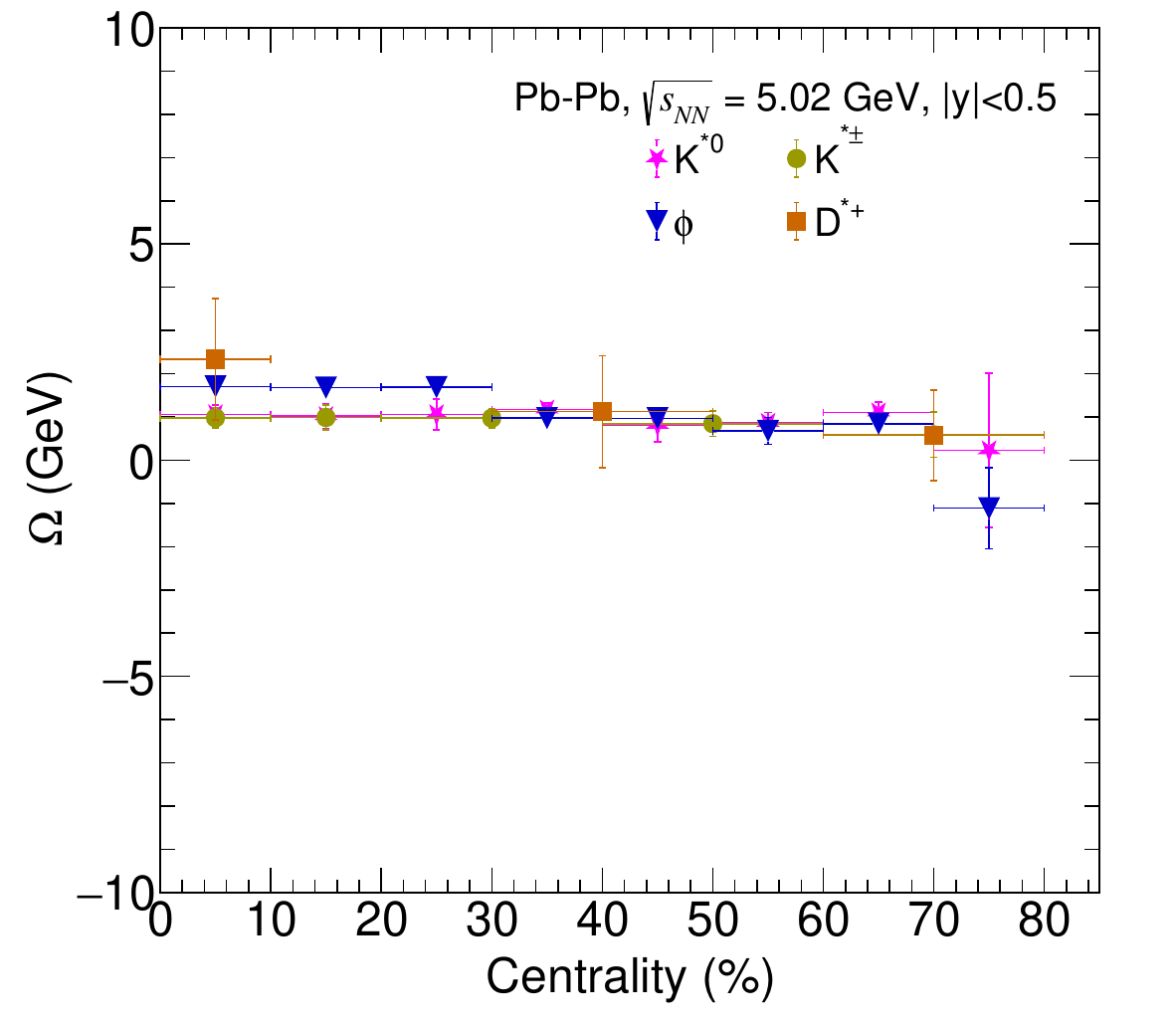}
\caption{The left and right panels show the variation of global vorticity $\Omega$ as a function of collision centrality for vector mesons obtained in Pb+Pb collisions at mid-rapidity for $\sqrt{s_{\rm NN}} =$ 2.76 TeV and 5.02 TeV, respectively.}
\label{Fig:ALICEMesons}
\end{figure*}
\subsection{Strange and Charm Vector Mesons}
Following a similar procedure to that for hyperons, we explore the rotational dependence of various vector mesons across centrality classes and a wide range of center-of-mass energies.
 
\subsubsection{Global vorticity at RHIC energies}
We analyze the $p_{\rm T}$-spectra of $K^{*0}$ and $\phi$ mesons measured by the STAR collaboration in Au+Au collisions at mid-rapidity ($|y|<0.5$) using the modified Tsallis distribution for various centrality classes and center-of-mass energies $\sqrt{s_{\rm NN}}=7.7$–39 GeV. The experimental data are taken from Ref.~\cite{STAR:2022sir}. Figure~\ref{Fig:STARMesons} presents the centrality dependence of the extracted global vorticity $\Omega$ for $K^{*0}$ and $\phi$ mesons. The results show that $\Omega$ exhibits a clear particle-species dependence across centrality classes, following a similar explanation as hyperons (anti-hyperons). Though both $K^{*0}$ and $\phi$ mesons are spin-1 particles, their rotational response differs, reflecting the influence of mass and quark content on spin–vorticity coupling, similar to the behavior observed for hyperons and anti-hyperons. In contrast to hyperons, it is observed that for most of the collision energies shown in Fig.~\ref{Fig:STARMesons}, the vorticity associated with $K^{*0}$ mesons is smaller in central collisions and increases toward peripheral events, whereas $\phi$ mesons display an opposite centrality trend. Therefore, it may be inferred that the rotational alignment may influence the spin alignment of $K^{*0}$ and $\phi$ mesons. These distinct patterns suggest that rotational effects at freeze-out play an important role in shaping the spin alignment of vector mesons and provide further evidence of particle-dependent coupling to the common vortical background of the medium.

\subsubsection{Global vorticity at LHC energies}
We further extend the analysis to LHC energies by studying the transverse momentum ($p_{\rm T}$) spectra of $K^{*0}$, $K^{*\pm}$, $\phi$, $\rho$, and $D^{*+}$ mesons produced in Pb+Pb collisions at mid-rapidity ($|y|<0.5$) and measured by the ALICE collaboration. The left panel of Fig.~\ref{Fig:ALICEMesons} presents the centrality dependence of the extracted global vorticity $\Omega$ for $K^{*0}$, $\phi$, and $\rho$ mesons at $\sqrt{s_{\rm NN}}=2.76$ TeV, using data from Refs.~\cite{ALICE:2014jbq, ALICE:2018qdv}. The magnitude of vorticity for $K^{*0}$ and $\phi$ mesons reaches its largest values in central collisions and gradually decreases toward peripheral events. The rotational behavior of the $\rho$ meson exhibits a weak dependence on centrality, suggesting a comparatively uniform response to the vortical medium. Furthermore, the right panel of Fig.~\ref{Fig:ALICEMesons} shows the corresponding results at $\sqrt{s_{\rm NN}}=5.02$ TeV for $K^{*0}$, $K^{*\pm}$, $\phi$, and $D^{*+}$ mesons, based on data taken from Refs.~\cite{ALICE:2021ptz, ALICE:2023ifn, ALICE:2018lyv}.  In the mid-central collisions (10-30)\% the value of $\Omega$ is higher for $\phi$ compared to  $K^{*0}$, $K^{*\pm}$. However, at peripheral collisions, i.e, (30-70)\% all the vector mesons $K^{*0}$, $K^{* \pm}$,  $\phi$, and $D^{*+}$ have a similar value of $\Omega$. It is evident at $\sqrt{s_{\rm NN}}=2.76$ TeV and $\sqrt{s_{\rm NN}}=5.02$ TeV as a small variation correspond to the chosen $p_{\rm T}$ fit range (0-5) GeV/c, and (0-20) GeV/c (at most central collisions), respectively. 

This study establishes a quantitative magnitude for the global vorticity generated in relativistic heavy-ion collisions, which can serve as a useful thermodynamic input for hydrodynamic simulations of QCD matter under rotation. The extracted vorticity offers valuable insights into the particle-species dependence of spin polarization and spin alignment. Different hadrons act as complementary probes of the rotational structure of the quark–gluon plasma: hyperons, which decouple earlier and carry larger intrinsic spin contributions, preserve pronounced imprints of the vortical field, while vector mesons reflect rotational effects shaped by later-stage hadronic interactions. Together, these observations enable a comprehensive characterization of the magnitude and evolution of vorticity throughout the system's lifetime.

\section{Summary}
\label{sum}

In this work, we have estimated the global vorticity generated in relativistic heavy-ion collisions by analyzing one of the key global observables: the transverse-momentum spectra of hyperons and vector mesons at RHIC and LHC energies.  The important findings of this paper and the future outlook are summarized below:

\begin{itemize}
\item  Global vorticity serves as a powerful probe for investigating how quark-gluon plasma responds to local vorticity, as examined in analytical and numerical studies of its thermodynamic properties.  We investigate how the strength of spin-vorticity coupling depends on particle mass, spin, charge, quark constituents (baryon vs meson), lifetime, hadronization mechanisms, and freeze-out conditions.

\item  We found that the global vorticity decreases towards peripheral collisions for $\Lambda$ and $\Xi^{-}$ hyperons at RHIC energies, whereas it remains independent of centrality at LHC energies. The $\Omega^{-}$ hyperons seem to have centrality dependence at both RHIC and LHC energies.

\item  The magnitude of the global vorticity produced at LHC energies is found to be maximum compared to RHIC energies. This analysis provides new insights into the behavior of QCD matter under rotation, particularly regarding the spin polarization of hyperons and the spin alignment of vector mesons arising from local vortical structures.

\item In addition to providing insight into the rotational dynamics of the QGP, global vorticity has far-reaching implications for the fundamental properties of the medium. It opens a novel avenue for exploring the QCD phase diagram, as vorticity can couple non-trivially to conserved charges such as baryon number, as well as to chiral dynamics and critical fluctuations.

\item  With recent developments in magneto-spin hydrodynamic frameworks for relativistic heavy-ion collisions, it has become particularly compelling to investigate the interplay between vorticity, shear-induced polarization, electromagnetic fields, and possible spin Hall–type effects, and their impact on a wide range of observables—an area that continues to stimulate intense theoretical activity.

\item With ongoing and upcoming experimental programs at RHIC, the LHC, NICA, FAIR, and future EIC facilities, the study of vorticity in the QGP is poised to become a central tool for uncovering fundamental aspects of strongly interacting matter under rotation.
    
\end{itemize}

\section*{Acknowledgment}
Bhagyarathi Sahoo acknowledges the Council of Scientific and Industrial Research (CSIR), Govt. of India, for financial support. The authors gratefully acknowledge the DAE-DST, Government of India, funding under the mega-science project "Indian Participation in the ALICE experiment at CERN" bearing Project No. SR/MF/PS-02/2021-IITI (E-37123).

\end{document}